\documentclass[journal]{vgtc}                     % final (journal style)

\usepackage{array}
\newcolumntype{L}[1]{>{\raggedright\arraybackslash}p{#1}}

\onlineid{0}

\vgtccategory{Research}

\title{Situatedness in Visualization Design: Making Unresolved Work Actionable}

\author{%
  \authororcid{Paul C. Parsons}{0000-0002-4179-9686},
  \authororcid{Prakash Shukla}{0009-0002-7416-1758},
  \authororcid{Phuong Bui}{0009-0004-3762-4770},
  \authororcid{Srishti Agrawal}{0009-0006-8745-6271},
  \authororcid{Ali Baigelenov}{0009-0003-6491-1874}
}

\authorfooter{
  \item
  	Paul Parsons is with Purdue University.
  	E-mail: parsonsp@purdue.edu
  \item
  	Prakash Shukla is with Purdue University and the University of Kansas.
  	E-mail: shukla@ku.edu.
    \item
  	Phuong Bui is with Purdue University.
  	E-mail: bui42@purdue.edu.
    \item
  	Srishti Agrawal is with Purdue University.
  	E-mail: agraw219@purdue.edu.
  \item Ali Baigelenov is with Purdue University and SDU University.
  	E-mail: ali.baigelenov@sdu.edu.kz.
}

\abstract{%
Visualization design often proceeds under unresolved conditions---goals shift, data remain provisional, stakeholder needs evolve, and several plausible directions may remain available at once. Existing visualization frameworks help organize design work and articulate major decisions, yet offer limited explanation of how practitioners proceed before a path forward has become clear. Drawing on an episode-level analysis of a previously collected three-phase qualitative corpus involving eleven expert visualization practitioners, we examine situations in which the problem, representational target, or viable direction remained unsettled. We find that practitioners make such situations actionable through provisional local moves. These moves reveal patterns, distinctions, and interpretive possibilities; clarify what is tractable, viable, or worth pursuing; and sometimes reorient the work itself. The analysis shows that situated action, professional judgment, and explicit design reasoning are intertwined in expert practice. It also identifies a practical limit on how fully design activity can be specified in advance. When the meaning of the next move depends on what a situation reveals in response to action, prescriptive decision structures cannot fully determine the course of design. The paper contributes an empirical account of situatedness in visualization practice and explains how local action makes unresolved work interpretable enough for consequential design decisions.
}


\graphicspath{{figs/}{figures/}{pictures/}{images/}{./}} % where to search for the images

\usepackage{tabu}                      % only used for the table example
\usepackage{booktabs}                  % only used for the table example
\usepackage{lipsum}                    % used to generate placeholder text
\usepackage{mwe}                       % used to generate placeholder figures
\usepackage{ccicons}                   % package to be able to use icons from creative commons

 \usepackage{array}
 \usepackage{tabularx}
 \usepackage[table]{xcolor}
 \usepackage{caption}

\newcolumntype{Y}{>{\raggedright\arraybackslash}X}

\usepackage{mathptmx}                  % use matching math font

\begin{document}

%%%%%%%%%%%%%%%%%%%%%%%%%%%%%%%%%%%%%%%%%%%%%%%%%%%%%%%%%%%%%%%%
%%%%%%%%%%%%%%%%%%%%%% START OF THE PAPER %%%%%%%%%%%%%%%%%%%%%%
%%%%%%%%%%%%%%%%%%%%%%%%%%%%%%%%%%%%%%%%%%%%%%%%%%%%%%%%%%%%%%%%

\firstsection{Introduction}

\maketitle

Data visualization design often reaches moments when the next step is not yet clear. Goals shift, data remain provisional, stakeholder needs evolve, and several plausible directions may remain available at once. In such situations, designers still have to move the work forward. They must determine what kind of situation they are in, what matters under those conditions, and what kind of move could make progress possible. Existing visualization frameworks provide resources for reasoning about design once a recognizable question can be posed---for example, what task to support, what encoding or interaction technique to choose, or what abstraction to use. They are less well suited to explaining how practitioners proceed before the work has stabilized enough for those decisions to become clearly meaningful.

This problem is salient in visualization design. Practitioners work with heterogeneous and often unstable datasets, evolving stakeholder goals, and tight coupling between data, representation, and interpretation. They frequently must move forward despite unresolved questions about data quality, framing, audience, and use context. Empirical studies of visualization practice suggest that progress in such settings depends less on following prescriptive sequences than on renegotiating goals, balancing trade-offs, and committing provisionally as conditions evolve \cite{parsons_understanding_2022, mckenna_design_2014, zhang_visualization_2023}. Making unsettled work actionable is a central part of visualization expertise. Without a clearer account of this aspect of practice, an important component of expert work remains difficult to study, teach, and support.

The need to act while goals, constraints, and possible directions remain unsettled is consistent with long-standing accounts of design as ill-structured, wicked, and indeterminate \cite{Cross1981, rittel_dilemmas_1973, buchanan_wicked_1992, stolterman_nature_2008}. Under such conditions, designers do not simply retrieve solutions from a fixed space of possibilities---they frame and reframe situations \cite{schon_problems_1984, Dorst_2015_FrameInnovation}, make tentative moves \cite{schon_designing_1988}, and allow problems and responses to develop together as they work \cite{maher_co-evolution_2003, cross_design_2011}. Open-ended design therefore often proceeds before a single interpretation of the work has stabilized.

In response to this gap in visualization, and drawing on broader theories of situated design action, we focus on \emph{situatedness} in visualization design practice. By situatedness, we mean more than design happening ``in context.'' We mean the local organization of design activity---how practitioners interpret what kind of situation they are in, what features become salient, what moves become plausible next, and how those moves reorganize the situation in turn. 

Existing visualization frameworks and empirical studies provide important starting points for this account. Frameworks help articulate decision loci, process activities, goals, tasks, problems, and design spaces \cite{munzner_nested_2009, meyer_nested_2015, mckenna_design_2014, lam_bridging_2018, schulz_design_2013, gleicher_problem_2023, dai_how_2026, sedig_design_2016, brehmer_multi-level_2013}. Empirical work has shown that visualization practice is opportunistic, contingent, and shaped by changing framings, constraints, and interpretations \cite{bigelow_reflections_2014, bigelow_iterating_2017, alspaugh_futzing_2019, walny_data_2020, zhang_visualization_2023, lee-robbins_client-designer_2025, parsons_beyond_2026, williams_data_2023}. What remains less fully explained is how local design activity is organized when the work remains unresolved, and how that activity becomes intertwined with more recognizable design decisions over time.

This perspective does not treat local action and explicit decision making as cleanly separate layers of design practice. Instead, it directs attention to how provisional moves, evolving constraints, representational considerations, and more recognizable design decisions interact in the flow of work. Our analysis draws on design cognition work that treats designing as situated activity \cite{chiu_design_2003, hay_future_2020} and emphasizes the sequential, locally consequential character of design action \cite{schon_designing_1992, schon_kinds_1992, christensen_kinds_2021}. Examining practice at this level helps clarify how progress is organized when the work has not yet stabilized around a clear next step.

To examine how practitioners make unresolved situations actionable, we present a new analysis of a previously collected three-phase qualitative corpus with eleven expert visualization practitioners, consisting of a shared design challenge, longitudinal design diaries from ongoing projects, and follow-up interviews. Here we focus on unresolved episodes and the situated organization of local activity within them, examining how practitioners make progress when the work remains unsettled and the next step is not fully clear.

Our study is guided by two research questions:
\begin{itemize}
  \item \textbf{RQ1:} How is situatedness expressed in episodes of unresolved visualization design work?
  \item \textbf{RQ2:} How do expert practitioners make progress when design situations are unresolved?

\end{itemize}

This paper contributes an empirical account of \emph{situatedness} in visualization design practice. We show how practitioners make unresolved situations more actionable through provisional action, and how such action can reveal interpretive possibilities, clarify tractability, and reorient the work itself. In doing so, the paper explicates a consequential dimension of visualization design that existing frameworks and practice accounts often rely on but do not fully elaborate. This perspective has implications for how researchers study practice, how educators prepare students for open-ended work, and how tools---including AI systems---might better support professional visualization activity in complex, data-rich settings.

\section{Background}

\subsection{Visualization Design Frameworks and Models}

Visualization research has developed a rich set of frameworks for reasoning about design work. These frameworks help articulate design decisions, process activities, and problem or design spaces, providing strong resources for describing, justifying, and evaluating visualization work at multiple levels.

Some frameworks focus primarily on \emph{design decisions} and their consequences. Munzner's Nested Model \cite{munzner_nested_2009} articulates four nested levels of visualization design and validation and links them to distinct threats such as wrong problem, wrong abstraction, wrong encoding, and wrong algorithm. Meyer et al.'s Nested Blocks and Guidelines extension \cite{meyer_nested_2015} supports finer-grained analysis and captures the rationale behind visualization design decisions. Other frameworks focus more directly on \emph{process activities}. McKenna et al.'s Design Activity Framework \cite{mckenna_design_2014} addresses the messy, iterative, and multilinear character of visualization work by connecting what designers do to the decisions they make across overlapping activities. Design Study Methodology \cite{sedlmair_design_2012} likewise provides an important methodological account of problem-driven visualization work, emphasizing that tasks may begin \emph{fuzzy} rather than crisp and that design-study work unfolds through iterative, overlapping stages rather than a rigid sequence. Still other frameworks support reasoning about \emph{goals, tasks, problems, and spaces}, including Lam et al.'s analysis goals \cite{lam_bridging_2018}, Schulz et al.'s design space of visualization tasks \cite{schulz_design_2013}, and accounts of problem and design spaces by Gleicher et al. \cite{gleicher_problem_2023} and Dai et al. \cite{dai_how_2026}.

Research has also moved toward more explicit accounts of visualization reasoning. Kindlmann and Scheidegger \cite{kindlmann_algebraic_2014}, for example, propose an algebraic model of visualization design based on relationships among data structure, representation, and perception. Wood et al.'s \cite{wood_design_2019} literate visualization work seeks to make design choices, assumptions, and reactions more visible as work unfolds, while Hutchinson et al. \cite{hutchinson_capturing_2025} treat design rationale itself as an object of analysis through question--answer--rationale triples derived from literate visualization notebooks. These efforts show growing interest in explicating visualization reasoning, including situated judgments and unresolved conditions that arise in practice. At the same time, their primary focus remains on making such reasoning more explicit once the relevant structures, tasks, or rationale can be articulated.

\subsection{Empirical Studies of Visualization Design Practice}

A growing body of visualization research has examined how designers work in real-world settings, moving closer to the contingencies of day-to-day practice. These studies show that visualization design does not typically unfold as a clean sequence from stable problem definition to finished solution. Instead, practitioners move opportunistically across tools, representations, and roles, often revising their understanding of the problem as they work. Early studies highlighted the importance of sketching, make-to-think activity, and the frictions of moving between drawing and programming environments \cite{bigelow_reflections_2014, bigelow_iterating_2017}. Subsequent work has shown that practitioners often proceed non-linearly when goals are unsettled, experimenting with partial artifacts, available tools, and emerging constraints rather than following a fixed path \cite{alspaugh_futzing_2019, walny_data_2020, zhang_visualization_2023, parsons_understanding_2022}.

This research also makes clear that practitioners do not begin from a single canonical foothold. Some start with available software, code, or data structures; others begin with narrative possibilities, communication goals, or stakeholder concerns; still others use sketches, examples, or precedent to discover what may be worth pursuing \cite{bigelow_reflections_2014, alspaugh_futzing_2019, bako_understanding_2023, baigelenov_how_2025, lee-robbins_client-designer_2025, parsons_understanding_2022}. Related work on data abstraction and mental models shows that even when designers confront the same dataset, they may prioritize different attributes, notice different structures, and connect the data to different abstract types \cite{williams_data_2023}. What becomes salient, and therefore actionable, depends not only on the data or brief itself, but on how designers locally construe what kind of situation they are facing.

Other studies show that visualization practice is shaped by negotiation with collaborators, clients, organizational demands, and shifting constraints. Designers often renegotiate scope, redefine goals, adapt to role boundaries, and work within tensions between analytical depth, communicative clarity, and production realities \cite{zhang_visualization_2023, ridley_sociocultural_2021, lee-robbins_client-designer_2025, parsons_judgment_2025}. Recent work has also emphasized the importance of judgment in these settings \cite{parsons_design_2020, parsons_fixation_2021, parsons_judgment_2025}, while related accounts of framing and problem--solution co-evolution show that progress often depends on reshaping the problem through action rather than solving a fully specified problem from the outset \cite{parsons_beyond_2026}. Methodologically oriented work likewise shows that visualization artifacts are shaped by people, context, disruption, and informal evaluation throughout the design process \cite{McCurdy2016}. Adjacent work on literate visualization and design-rationale capture also suggests that practitioners' justifications, alternatives, and evolving rationales are worth documenting and studying, even if such work has focused more on explicit exposition than on the situated organization of action in unresolved episodes \cite{wood_design_2019, hutchinson_capturing_2025}.

%These studies establish that visualization design practice is contingent, adaptive, and shaped by local material, organizational, and interpretive conditions. What remains less developed is an explicit account of how practitioners make such contingency actionable in local episodes of work. It is this aspect of practice that our paper seeks to describe more explicitly.

\subsection{Design Cognition and Local Design Action}

To clarify this underdescribed layer of visualization practice, we draw on design cognition research that treats designing as situated, locally organized action. From this perspective, design depends not only on search, but also on framing, registration, and ongoing engagement with local representations and materials \cite{suchman_plans_1987, Kirsh2009, schon_reflective_1983}. Action is therefore not merely the execution of prior reasoning; it is part of how a situation becomes interpretable enough for further reasoning to proceed. More recent work in design cognition extends this orientation by emphasizing situated cognition, temporality, and context-sensitive activity in naturalistic settings \cite{hay_future_2020}. Recent work in visualization characterizes this situated coordination as knowledge-in-use---the interpretation, adaptation, and coordination of multiple forms of knowing in concrete design situations \cite{parsons_design_2027}.

This orientation is important here because it provides a way to study visualization design at a more local analytic grain than project-level process descriptions alone. Chiu \cite{chiu_design_2003} argues that situated studies focus on how behavior and cognition are shaped in relation to specific situations, and treats design as a situated activity at the micro level in which designers interact with cases, conditions, and tools rather than simply applying abstract rules. This perspective is helpful for our purposes because it shifts attention away from whether a project has entered a given stage and toward how a designer interprets a local situation, what features become salient, and what forms of action become plausible under those conditions. In other words, it supports treating situatedness not as a generic synonym for context, but as the local organization of design activity.

Classic design scholarship also points toward this level of analysis. In Schön's account, design proceeds through a reflective conversation with the situation, often described as ``see--move--see''---the designer reads what is there, makes a move, and then reads the consequences of that move in light of an evolving appreciation of the situation \cite{schon_designing_1992, schon_kinds_1992}. This account captures an important feature of unresolved work, in that progress depends on provisional action that changes what the designer is able to notice next. Christensen and Abildgaard similarly argue that design progression can be understood through the sequential ordering of local actions and their consequences, rather than only through high-level process descriptions \cite{christensen_kinds_2021}. Their work shows that local design activity may be organized through movements that are materially and temporally situated, even when those movements do not map neatly onto named methods or stages.

\section{Methods}

\subsection{Corpus and Relationship to Prior Publications}

This paper presents a new analysis of a previously collected multi-phase qualitative corpus on expert visualization design practice. The corpus has informed prior publications from this research program, including our earlier paper on framing and problem–solution co-evolution in visualization design \cite{parsons_beyond_2026}. The two papers examine the same broader corpus at different analytic grains and for different purposes.

The earlier paper focused on project-level development, asking how practitioners frame design situations and how problems and solutions co-evolve over the course of a project. By contrast, this paper focuses on bounded unresolved episodes within and across projects, asking how local design activity is organized when practitioners encounter situations that are not yet sufficiently settled to make the next step clear. In this sense, the current analysis shifts attention from how projects evolve through framing and co-evolution over time to how practitioners act within specific unresolved moments to make the work more actionable.

This distinction also explains why some empirical episodes appear in both papers. The same episode may be analytically relevant at more than one level. In the earlier paper, such episodes were used to show how framing and problem--solution relations changed across a project. Here, they are used to examine how practitioners interpret unresolved situations, make provisional moves, and respond to what those moves reveal. The contribution of the present paper is therefore a more local account of situated action within unresolved design work.

We draw on a three-phase qualitative corpus consisting of (1) a shared, time-boxed design challenge; (2) a longitudinal diary study embedded in participants' ongoing professional projects; and (3) follow-up semi-structured interviews. These three phases were designed to provide complementary views of practice. The challenge created a common prompt that made cross-participant comparison possible. The diary study captured in-flight work as it unfolded in real projects. The interviews supported retrospective walkthrough and cross-phase reflection. All three phases involved the same participants.

\begin{table}[t]
\centering
\small
\setlength{\tabcolsep}{3pt}
\renewcommand{\arraystretch}{1.08}
\caption{Comparison with prior project-level analysis \cite{parsons_beyond_2026}.}
\label{tab:relation-prior-work}
\begin{tabular}{p{0.16\columnwidth} p{0.38\columnwidth} p{0.38\columnwidth}}
\toprule
 & \textbf{Prior paper \cite{parsons_beyond_2026}} & \textbf{This paper} \\
\midrule
\textbf{Focus} 
& Framing and problem--solution co-evolution 
& Situated action in unresolved episodes \\

\textbf{Unit} 
& Project trajectories 
& Bounded unresolved episodes \\

\textbf{Question} 
& How do problems and solutions co-evolve across projects? 
& How do practitioners proceed when the next step is unclear? \\

\textbf{Contribution} 
& Explains project development through framing and co-evolution 
& Explains how provisional action makes unresolved situations interpretable, tractable, and reoriented \\
\bottomrule
\end{tabular}
\end{table}

\subsection{Participants}

We purposefully sampled expert practitioners rather than novices because our interest was in experienced judgment and situated action in professional visualization work. To identify participants, we drew on Hoffman's Pentapod Principle \cite{hoffman_identifying_2023} and prior scholarship on design expertise \cite{atman_engineering_2007}, treating expertise as multifaceted rather than reducible to years of experience alone. Recruitment therefore emphasized visible markers of expertise---for example, leadership on visualization teams, awards, invited talks, authorship, exhibitions, and peer recognition.

We recruited through professional networks and community channels, including mailing lists, practitioner groups, and social media. Eleven practitioners participated across domains including newsrooms, agencies, product teams, and independent practice. Pseudonyms are used throughout. Participants were recruited from March 2023 through February 2024, and data were collected from September 2023 through July 2024. The study was reviewed and approved by the Institutional Review Board at Purdue University under protocol no. 2022-1273, and all participants provided informed consent prior to participation. Participants received compensation of \$250 USD. Table~\ref{tab:participantdemographics} provides an overview of the 11 participants, including pseudonyms, self-reported titles, years of experience, and visible markers of expertise.

\begin{table}[t]
\centering
\small
\setlength{\tabcolsep}{4pt}
\renewcommand{\arraystretch}{1.1}
\caption{Demographics of study participants (pseudonyms, titles, experience, and expertise indicators).}
\label{tab:participantdemographics}
\begin{tabularx}{\columnwidth}{l Y c Y}
\toprule
Pseudonym & Title & Exp.\ (yrs) & Expertise Indicators \\
\midrule
John   & Data Experience Designer      & 11+   & Speaker; awards; peer recognition \\
Priya  & Data Visualization Specialist & 2--5  & Awards; peer recognition \\
Elena  & Lead Visualization Designer   & 6--10 & Speaker; leads visualization team \\
Carlos & Data Visualization Designer   & 6--10 & Speaker; awards; exhibitions \\
Omar   & Senior Interactive Designer   & 11+   & Speaker; peer recognition; team lead \\
Lauren & Creative Information Designer & 11+   & Speaker; awards; peer recognition; author \\
Jade   & Data Visualization Designer   & 6--10 & Speaker; peer recognition \\
Lina   & Data Visualization Designer   & 6--10 & Awards \\
Sofia  & Information Designer          & 2--5  & Speaker; awards; exhibitions \\
Maya   & Data Visualization Specialist & 11+   & Speaker; awards; peer recognition \\
Laila  & Visual Data Journalist        & 6--10 & Awards; peer recognition \\
\bottomrule
\end{tabularx}
\end{table}

\subsection{Phases, Data, and Materials}

Across phases we collected probe reflections, diary entries, interviews, and associated artifacts such as sketches, screenshots, and intermediate visualizations. We treated these artifacts as analytic materials because they often captured provisional moves, failed attempts, and local reorientation later discussed in reflection and interview.

In the first phase, each participant completed an independent, time-boxed design probe (suggested duration: 90 minutes). We based the probe on the inaugural ``Data Is Plural'' design challenge hosted by the Data Visualization Society in 2022 \cite{nightingale_editors_data_2022}. The challenge uses a publicly available, multi-table historical dataset and an intentionally open brief, placing all participants in a comparable unresolved situation without prescribing a target question, audience, or visual direction. This phase therefore provided a common point of comparison for examining how practitioners initially established footholds and proceeded when the path forward was similarly open.

Participants received the original public brief and dataset along with instructions emphasizing process and reasoning rather than polished output. Each participant provided two written reflections, one immediately after first examining the brief and data, and one after completing the probe. These reflections asked what initially stood out, what felt unresolved, what they tried, what changed, and what they would do next with more time. Participants could also attach process artifacts. In the second phase, participants kept a reflective diary while working on an active professional visualization project for up to eight weeks. Each selected an ongoing project they expected to continue over several weeks, such as a newsroom piece, client deliverable, or product feature. We asked for approximately one entry per week, with flexibility to match project rhythms. Diary prompts focused on unresolvedness and action in practice, including what felt unsettled, what options were on the table, what they decided to do next, and whether anything had led them to reconsider their framing, audience, priorities, or scope. In the third phase, each participant completed a remote semi-structured interview over Zoom, averaging roughly 90 minutes. Interviews were recorded, automatically transcribed, and manually checked for accuracy. We revisited the design-challenge materials and then returned to the diary project using entries and artifacts as prompts to discuss turning points, constraints, stakeholder dynamics, and episodes in which several paths were possible. We also asked broader questions about how participants typically respond when a project feels unresolved or stuck.

In total, the corpus comprises 22 probe reflections, 11 diary sequences spanning up to eight weeks each (median ~7 entries per participant, ~330 words per entry), and approximately 16 hours of interview recordings (11 sessions, 72--101 minutes, average ~90 minutes), plus dozens of sketches, screenshots, and intermediate visualizations.

\subsection{Unresolved Episodes as the Unit of Analysis}

The primary unit of analysis in this paper is the \emph{episode}, rather than the project or participant as a whole. We define an unresolved episode as a bounded segment of practice or reflection in which a participant encountered a situation that was not yet sufficiently settled to make the next step clear and had to determine how to proceed. Such situations included, for example, unclear scope, competing priorities, unstable or weakly understood data, representational uncertainty, stakeholder misalignment, shifting requirements, and time or production pressures.

This episode-level focus distinguishes the present analysis from earlier project-level analyses of the same corpus. Rather than examining how framing and problem--solution relations developed across a whole project, we focus here on how practitioners locally interpreted what kind of situation they were in, what they treated as salient, what provisional moves they made, and how the consequences of those moves changed what became possible next.

\subsection{Analysis}

We conducted a reflexive thematic analysis focused on the situated organization of local design activity. Our approach combined deductive and inductive analysis \cite{braun_using_2006, fereday_demonstrating_2006, swain_hybrid_2018}. We began with sensitizing concepts from visualization research and design scholarship---including framing, problem--solution co-evolution, reflection-in-action, judgment, abstraction, and representation---but allowed the analysis to be shaped by participants' own accounts of unresolved situations and how they responded to them.

Analysis proceeded iteratively rather than through a strictly linear coding pipeline. We first read across the full corpus to identify episodes in which participants described or enacted locally unresolved situations and some attempt to move the work forward. Episodes were marked when there was evidence of both (1) an unsettled condition and (2) an interpretive or practical response, such as testing a representation, narrowing scope, restructuring data, revising a question, or reconsidering feasibility. Probe reflections, diary entries, interview walkthroughs, and submitted artifacts were treated as mutually informative in identifying and reconstructing these episodes.

We then compared episodes within and across participants and phases, attending to three analytic questions: what made the situation unresolved, what kind of provisional action was taken, and what consequence that action had for the direction of the work. Through repeated comparison, we refined both the boundaries of unresolved situations and the distinctions among consequences of action, arriving at the recurrent patterns reported below---unresolved situations, provisional action as discovery or tractability-making work, and reorientation as a consequence of acting within such situations.

We moved repeatedly between individual episodes, participant-level context, and cross-case patterns. We used Dovetail to manage transcripts, diary entries, reflections, and associated artifacts. Four authors reviewed stratified subsets of the corpus spanning different participants and phases, then introduced overlap to compare interpretations, challenge emerging distinctions, and refine analytic boundaries. Coding remained interpretive and evolving throughout this process. We developed codes iteratively and continually revisited, refined, and contested them as we moved across episodes and phases. We did not use a fixed codebook to standardize coding across analysts. Consistent with reflexive thematic analysis, overlapping review was used to surface alternative readings and deepen the developing analysis; coder agreement was not treated as a criterion of validity, and we did not compute inter-coder reliability \cite{mcdonald_reliability_2019}.

Our aim was not to exhaustively catalog all possible behaviors, nor to quantify their frequency, but to develop a grounded account of how expert visualization practitioners make unresolved situations more actionable in day-to-day design work. Across the corpus, we identified and analyzed 35 unresolved episodes spanning the design challenge, diary, and interview phases. The patterns reported below did not appear uniformly across all participants or episodes; rather, they recurred across challenge, diary, and interview materials, with different episodes foregrounding different unresolved situations, provisional actions, and consequences.

\section{Findings}
Our analysis suggests that unresolved situations in visualization design practice do not become actionable simply through deliberation or by applying predefined methods. Across the corpus, practitioners encountered situations in which what mattered, what was possible, and what should happen next were not yet clearly settled. In response, they made provisional moves that helped make the situation more actionable. These moves did not all do the same kind of work. In some cases, action revealed a new pattern, distinction, or interpretive possibility. In others, it clarified tractability by showing what was too messy, too costly, too constrained, or too weak to pursue further. Across both forms, action was not merely a response to an already understood problem; it was part of how the next workable step became available. Some of the episodes below come from the shared design challenge, which was especially useful for comparison because all participants began from the same open brief and dataset but established footholds in different ways. Table~\ref{tab:episode-summary} summarizes these findings using the same terms developed below and identifies the sections in which each recurrent situation is discussed.

\begin{table}[ht]
\centering
\small
\setlength{\tabcolsep}{3pt}
\renewcommand{\arraystretch}{1.08}
\caption{Summary of recurrent unresolved situations, typical actions, and consequences across analyzed episodes.}
\label{tab:episode-summary}
\begin{tabular}{>{\raggedright\arraybackslash}p{0.22\columnwidth} >{\raggedright\arraybackslash}p{0.24\columnwidth} >{\raggedright\arraybackslash}p{0.25\columnwidth} >{\raggedright\arraybackslash}p{0.15\columnwidth}}
\toprule
\textbf{Situation} & \textbf{Unresolved aspect} & \textbf{Typical actions} & \textbf{Consequence} \\
\midrule
Underdetermined brief (\S\ref{sec:unresolvedsituations})
& Scope, question, threshold, narrative target
& Narrowing, comparing angles, testing partial structures
& Foothold \\
\addlinespace[2pt]
Unstable material (\S\ref{sec:unresolvedsituations})
& Missing or unreliable data; unstable mappings
& Restructuring, anomaly checking, bounding reliability, testing encodings
& Clarified tractability \\
\addlinespace[2pt]
Shifting conditions (\S\ref{sec:unresolvedsituations}; \S\ref{sec:reorientation})
& Changing requirements, misalignment, technical constraints
& Workarounds, fallback, negotiation, redesign
& Reorientation \\
\addlinespace[2pt]
Representational uncertainty (\S\ref{sec:actiondiscovery}; \S\ref{sec:reorientation})
& No clear visual structure or comparison logic
& Sketching, testing forms, reorganizing, simplifying
& Discovery / reorientation \\
\bottomrule
\end{tabular}
\end{table}

\subsection{Unresolved Situations in Visualization Design Practice}
\label{sec:unresolvedsituations}

Across the corpus, participants often encountered situations that had not yet settled into anything clearly actionable. These situations took four recurrent forms: underdetermined briefs, unstable material, shifting conditions, and representational uncertainty (see Table~\ref{tab:episode-summary}). These unresolved situations were not confined to the deliberately open-ended design challenge; they also appeared throughout participants' diary entries and interviews about ongoing professional work. Their significance lay in insufficient stability to determine the next step, beyond the ordinary messiness or iteration of design work. A more stabilized situation, by contrast, would have had the relevant problem, representational target, or project boundary sufficiently settled for a recognizable design or implementation decision to guide the work forward.

In some cases, unresolvedness took the form of an \textit{underdetermined brief}. In diary entries about his ongoing professional work, Carlos described a journalistic investigation into streaming fraud as the ``most difficult project'' he worked on that week precisely because there was no clear brief or predefined goal to work from. There were many possible data sources, many possible analytical angles, and no obvious threshold for deciding what would count as enough. As he put it, he was ``in a position to decide for myself how much is enough and what data analysis angle is more relevant,'' adding that the project would have been much easier if there had been ``a clear brief and a clear goal,'' but instead he had to ``be briefing [him]self, in a way.'' That uncertainty was compounded by the tension between classical journalistic work and exploratory data work, as there was always more data he could pursue, but no external structure telling him when the angle had become focused enough to justify moving forward. In a related project, discussed in his interview, he described trying to compress a story of Indigenous persecution spanning centuries into a single map without either overloading the piece or stripping away too much of what made the story matter. In both cases, the difficulty lay in determining what kind of visual story could be made coherent enough to pursue, rather than in choosing among already intelligible options. For example, Carlos had to determine what form of account the data could support, what could be made legible within a single visual frame, and what level of reduction would still preserve the significance of the story. It was not only a preferred direction that remained unavailable; rather, what was missing was a stable criterion for deciding which angle was relevant enough, complete enough, and representationally coherent enough to justify further work. A more stabilized situation would have been one in which a clear investigative angle, workable evidentiary threshold, and coherent visual narrative direction were already available.

A second recurrent form was \emph{unstable material}---underlying data, mappings, or other evidentiary material had not yet become stable enough to support a trustworthy visualization. In diary entries about a map-based civic technology visualization already nearing release, Lina described discovering counties in which the year-over-year changes in polling locations simply ``didn't seem right.'' The project used circles on a map to encode two variables at once---the number of people per polling location through size, and the change since the last election through color intensity. Under those conditions, the problem did not remain hidden in the background. Lina wrote that the visualization was becoming ``potentially misleading and/or confusing to users given how the encodings work.'' In fact, the very perceptual choices meant to make larger changes more legible were amplifying the instability. She had chosen ``perceptually deeper colors'' to represent larger increases or decreases, ``but that has the added effect of highlighting egregious data quality issues.’’ In a later entry, she described the consequence even more plainly---because the project visualized openings and closures over time, missing records could make it look as though ``a bunch of locations have either opened or closed because the records are missing for certain years.'' Rather than continuing with visual refinement, she and the team had to step back, assess the reliability of the affected areas, and temporarily remove them from the map. What made this situation unresolved was not merely that the data needed cleaning. It was that the act of visualizing had exposed a more basic problem of the represented situation not yet being stable enough to support a trustworthy public-facing account. The visualization’s encoding logic made the instability consequential for what users would take the situation to be, extending the problem beyond encoding selection for a known finding. A more stabilized situation here would have been one in which the mapped changes could be treated as reliable enough for the encodings to communicate a public-facing account rather than expose unresolved instability in the underlying records.

A third recurrent form was \emph{shifting conditions}: evolving requirements, stakeholder misalignment, or technical constraints changed while the work was underway. In diary entries about an ongoing dashboard project, Jade described how instability arose from a solution that kept slipping out from under the team as the work progressed. In one case, a change in requirements arrived ``in the middle of the project'' and forced a rebuild of the data-source architecture on which the dashboard depended. Because the needed data were no longer available in the expected form, she had to redirect her developers away from building the new dashboard and toward enhancing and formatting dashboards that had already been created. As she put it, implementing the new requirement was simply ``impossible,'' so the team had to negotiate with the business about what the final solution should actually look like, setting new boundaries around what kinds of synchronized refreshes were technically possible. The situation became still more unstable when a later accumulation of change requests made the original design impossible to implement at all. Jade realized this only after assigning one of her team members what she later called ``an impossible task,'' which then required her to step back into implementation work herself, look for a workaround, and de-prioritize other tasks. Across these entries, the problem was that the dashboard itself had not settled into a stable enough representational target for execution to proceed coherently. What could still be shown together, refreshed together, and meaningfully presented as a single system was no longer clear. That is why one of her most difficult decisions became ``deciding when to stop looking for the workaround solution and change the design.'' At that point, unresolvedness was no longer a matter of refining a chosen direction, but of determining what version of the dashboard remained feasible and representationally coherent at all. The issue concerned what could still count as a workable visual solution, requiring more than a revised implementation plan. %A more stabilized situation here would have been one in which the target functionality, technical constraints, and representational boundaries were settled enough for implementation to proceed without redefinition of the dashboard itself.

A fourth recurrent form was \emph{representational uncertainty}: no clear visual structure or comparison logic was yet available to organize the work. Because practitioners often addressed this condition by sketching, testing, and reorganizing partial representations, it becomes especially visible in the accounts of action as discovery (Section~\ref{sec:actiondiscovery}) and reorientation (Section~\ref{sec:reorientation}) below.

These episodes show that participants were not making decisions inside already intelligible problem spaces. They were working in situations whose scope, reliability, relevance, or direction had not yet stabilized enough to make the next step clear. The operative problem, representational target, and feasible project boundary themselves remained unresolved; participants were not choosing among already intelligible options. Existing models and frameworks remain valuable for articulating explicit decisions and process activities, but these episodes point to a less fully elaborated aspect of practice in which designers make provisional moves, see what those moves disclose, and continue in light of the reorganized situation.

\subsection{Provisional Action Makes Situations Actionable}
\label{sec:provisionalaction}

When participants encountered situations that were not yet stable enough to make the next step clear, they did not simply wait for the right interpretation to arrive. Instead, they acted into the situation. They pulled rough subsets, sketched provisional layouts, tested chart forms, restructured tables, grouped categories, and pushed partial representations just far enough to see what they would reveal or resist. These moves were often tentative, incomplete, and sometimes visibly unsuccessful. Participants were not only trying to externalize a settled idea. They were using visual and analytic action to probe the situation in which they were operating, determining what it could support and how the work might proceed from there.

Across the corpus, these provisional moves seemed to do at least two recurring kinds of work. In some episodes, they exposed a new pattern, distinction, or interpretive possibility that had not been clear beforehand. In others, they clarified tractability by showing what was manageable, viable, or worth pursuing under the circumstances. In both cases, action was not merely downstream from understanding. It was part of how the work became more interpretable and actionable as it unfolded. %In a more settled situation, by contrast, practitioners could often move more directly into refining or implementing a viable direction already in view. In the episodes below, provisional action was needed to help establish what kind of direction, if any, the situation could support in the first place.

\subsubsection{Action as Discovery}
\label{sec:actiondiscovery}

In some episodes, provisional action functioned as discovery. Here, a move exposes something in the material that is not yet visible, while also testing whether a known idea can be executed cleanly. John's account was the clearest instance in our corpus. In an interview about a commissioned visualization on food pairings, he described being asked to rethink an existing force-directed network that showed similarities among foods based on shared chemical compounds. The problem was not simply to redraw the scientists' network more elegantly. He needed a different visual approach that would still communicate the same underlying relationships while remaining interesting and readable, even though the original network was dense, highly connected, and full of overlapping lines.

His first move was to reorganize the foods by category---fruits, vegetables, meats, and so on---and place those groups in rows. He then began drawing the compound-based connections between foods, only to find that the result was still too tangled to make much sense of. Rather than treat that failed arrangement as a dead end, he kept working the representation itself. He tried different line treatments, adjusted thickness and curvature, and eventually rotated the grouped foods from rows into columns to see whether the structure would hold together more clearly. It was during that representational work, not prior to it, that a more consequential distinction began to emerge. Some links connected foods \emph{within} the same category, while others connected foods \emph{between} categories, and the latter turned out to be the more surprising and interesting ones. As John put it, this was ``something I only figured out during this process,'' after which he began coloring the two types of connections differently. The move produced more than a more workable layout. A more settled situation would have been one in which the distinction organizing the design was already clear enough to guide representational choices from the outset. Instead, the move revealed a distinction in the data that became central to the final visualization itself.

Omar described a related logic, but from a different starting point. Rather than beginning with a fixed story he was trying to confirm, he deliberately adopted what he called a ``naive'' or inductive methodology, which referred to taking the dataset as if he knew nothing about it and first exploring its internal structure rather than its external meaning. He restructured the data in JavaScript, generated distributions and scatterplots, and tested relationships among categories such as authors, theaters, and time in order to see what kinds of patterns the material itself might disclose. Some expected paths did not hold strongly---for example, cross-category relationships among authors and theaters---while temporal patterns proved more revealing. In his case, provisional action did not refine an interpretation already in hand. It helped determine what kind of visual direction the data could plausibly sustain in the first place.

\subsubsection{Action as Tractability}
\label{sec:actiontractability}

In many of the other episodes we analyzed, however, provisional action clarified tractability rather than yielding discovery in this stronger sense. In these episodes, the move helped determine what was manageable, viable, supportable, or worth pursuing under the circumstances. The move established what the work could realistically sustain, as discovering what the data meant was only part of its function.

Priya's challenge work offers a clear example. Faced with a dataset that was too large to handle comfortably as a whole, she did not try to preserve the full space of possibilities. Instead, she actively searched for a subset that was, in her words, ``not too small, not too big and interesting enough.'' One candidate slice was too limited to do much with; another was too large to render reliably within her usual workflow. The immediate task was to find a portion of the dataset that could sustain a workable visual overview; selection of an ideal visual form came later. Once she had identified a manageable subset, she could test a simple chronological arrangement that was easy to build and quick to read. The chronological arrangement was technically feasible, preserved the individual entries as visible marks, and gave a general picture of how performances unfolded over time. She briefly considered a more geographical organization, but treated it as riskier and less effective for achieving that kind of overview. The move therefore did not reveal a rich conceptual breakthrough. A more settled situation would have been one in which the data and the representational target were already manageable enough for the work to move directly into refining or comparing alternatives. Instead, it created a foothold by establishing a subset and visual structure that the data could plausibly sustain.

Sofia's challenge account makes the same point in a more explicitly visual way. She first tried to work through actors and their relationships to performances, but the structure quickly became too tangled, with primary, secondary, and recurring roles making the relational picture hard to manage. Under time pressure, she narrowed toward performances and theaters and then tried an alluvial diagram as a way of getting something to show. But she did not treat the alluvial as a final answer. She treated it as a test---could this format hold the relationships in a way that remained legible? When the many-to-many structure and title inconsistencies made the attempt break down, the failed chart still did useful work. It clarified what the data and the deadline could bear. This is where she offered one of the clearest formulations in the corpus: ``I think I have to see things before I make [decisions] . . . I use visualization as a way to make mistakes.'' The value of the move lay precisely in making those mistakes early enough to redirect the work.

Laila's challenge work shows a similar dynamic, but through representational fallback. She began with more ambitious possibilities, including a heatmap-like direction and other chart forms that she explored through tools such as RAWGraphs. But those directions began to resist the shape of the data and the available time. Rather than continue pushing them in the abstract, she turned toward a simpler treemap-like structure built around the titles with the most performances, something she knew she could actually make legible and complete. Even within that simpler form, she continued adjusting scale, grouping, and placement, using the layout itself to see where proportion and readability were still not working. The move here did not reveal a new conceptual pattern in the data. A more settled situation would have been one in which the representational target was already clear enough for the work to focus on improving readability within a viable form. Instead, it clarified what kind of representation could be made to work at all under the conditions.

Lauren described a comparable pattern of productive failure. Her first exploratory question was whether one theater might have acted as a trend-setter, bringing performances to the public before other theaters caught on. After bringing the data into Tableau and trying to pursue that idea, however, she found that the pattern she was hoping for was simply not there in any compelling way. That failed path still proved useful. It redirected her toward other questions the data could better sustain, and it changed which visual forms now seemed plausible. A bar-chart logic gave way to a bump-chart logic once she realized that raw magnitude across theaters was misleading because the theaters differed so much in scale. But even that move generated a further practical problem, in that the connected lines began to read too much like time-series data, which was not what she wanted to communicate. Provisional action did not merely narrow the work in a superficial sense. A more settled situation would have been one in which the comparison question itself had already stabilized, allowing chart choice to function as a more familiar representational decision. Instead, the move clarified what was misleading, what was feasible, and where the next workable question might lie.

These episodes show that provisional action often made progress possible by clarifying tractability even when it revealed no striking new insight. A tested subset, a failed alluvial diagram, a fallback treemap, or a discarded trend-setter hypothesis could all do important work by showing what was too tangled, too costly, too misleading, or too ambitious to sustain. In that sense, action did not merely implement a direction that was already fully settled. It helped establish what kind of direction could be understood as workable in the first place.

\subsection{Reorientation as a Consequence of Provisional Action}
\label{sec:reorientation}

What provisional action revealed often reoriented the work itself instead of prompting another round of iteration within the same working understanding of the problem. As participants tested partial views, tried encodings, narrowed scope, or pushed on failing solutions, they were not only learning whether a given direction could be improved---they were often learning that the direction itself needed to change. In that sense, these episodes were not merely additional passes on the same provisional line of work. What shifted was the object of attention itself---the participant came to see a different problem, a different representational target, or a different boundary on what the project could plausibly sustain. In a more settled situation, participants could have revised or refined a viable direction while preserving the same underlying problem or representational target. In the episodes below, provisional action instead changed the direction itself.

Maya described this especially explicitly. Reflecting on the challenge, she explained that she rarely expects her first idea to be the best one and instead deliberately works through multiple candidate directions before deciding what is worth pursuing. To do this, she pushes herself to generate several distinct ways of structuring the same material in quick succession, using each attempt not simply as a stylistic alternative but as a way of testing what kind of \emph{visual framing} the dataset might actually support. As she described it, she initially did not yet have her core question and had to work through whether the relevant focus was the timeline, the theaters, the plays, the actors, or some broader historical impact. In that process, she not only explored multiple ideas but also rejected several representational directions. The verbatim-based route proved too incomplete to sustain, conventional charts were not strong enough to carry the story, and the dashboard format, while attainable, did not seem capable of making the material sufficiently interesting or meaningful. She ultimately moved toward a narrower four-theater structure that could support a more coherent historical comparison. The shift here was therefore not just iterative improvement to an already recognized design problem. It was a change in what kind of visualization problem she understood herself to be solving, and what kind of representational target the data could plausibly sustain. A more settled situation would have been one in which the relevant story structure and visual framing were already clear enough for her to refine a promising direction rather than move among competing ways of holding the material.

Sofia's challenge work shows a more compressed but equally important form of reorientation. She began by trying to organize the data around actors and their relationships to performances, but that path quickly became too tangled to sustain. Primary, secondary, and recurring roles made the actor-centered structure difficult to manage, and the more she pushed on it, the clearer it became that this was not simply a chart that needed refinement. It was no longer proving to be a workable way of holding the problem. She then narrowed toward performances and theaters and tried an alluvial diagram as a way of seeing whether that reduced structure could remain legible. When that route also began to break down under title inconsistencies and many-to-many complexity, the work still moved forward, but in a different direction. These were not just iterative attempts to perfect the same representational idea. The object of attention had changed, from actor relationships, to performance-theater relationships, to whatever the data and deadline could still support.

Jade's diary entries show reorientation under even more concrete project pressure. In her case, the thing being redirected was not a sketch or exploratory chart, but a dashboard project already underway. One requirements change arrived in the middle of implementation and forced a rebuild of the data-source architecture that the dashboard depended on. Later, accumulated requests pushed the design beyond what the existing implementation could support, a shift she only fully recognized after assigning what she later called ``an impossible task'' to one of her team members. At that point, the issue was no longer whether the team could keep refining the current solution. The object of work itself had changed. What had originally been a dashboard-design task had become a question of whether the project could still hold together in its current form at all. Reorientation here meant renegotiating scope, abandoning assumptions about what would be delivered, and deciding when to stop pursuing a workaround and change the design itself. That is not well described as iteration in the narrow sense of repeated refinement. It is a shift in what version of the project remains feasible enough to continue.

Lauren and Laila show related but quieter forms of reorientation. Lauren’s initial question was whether one theater might have acted as a trend-setter, bringing performances to the public before other theaters caught on. Once she began testing that idea in Tableau, however, the pattern she was hoping for did not materialize in a compelling way. That failure did not simply narrow the analysis; it changed the question she pursued and, with it, the visual forms that now made sense. Laila likewise began with more ambitious possibilities, including a heatmap-like route, but as those possibilities resisted both the data and the available time, she turned toward a simpler treemap-like structure she could actually make legible. In both cases, reorientation involved more than replacing one chart with another. It involved redefining what kind of representational problem the work could plausibly sustain.

These episodes suggest that reorientation should not be reduced to iteration in the familiar sense of revising or refining an existing direction. They are also related to what prior work has described as reframing \cite{parsons_beyond_2026}, although they are not entirely the same. The difference is one of analytic emphasis. Reframing refers to a broader shift in how the problem situation is understood. Reorientation, as used here, refers to the more local redirection of work that follows when provisional action changes what now appears workable, relevant, or worth pursuing. Some reorientations are indeed driven by reframing. Others are less extensive, involving a change in route without a stronger reformulation of the underlying problem—for example, when a designer drops one representational path, narrows the feasible scope, or adopts a fallback that the situation can still sustain. Rather than tracing how projects are reframed over time, we focus here on how provisional action within unresolved episodes redirects the work by altering what the situation is taken to require next.

\section{Discussion}

This paper contributes an empirical account of how unresolved situations become more actionable in visualization design practice. In answer to RQ1, the findings show that situatedness is expressed through bounded unresolved episodes in which practitioners must locally interpret what kind of situation they are in, what features matter under those conditions, and what kinds of moves are plausible next. In answer to RQ2, they show that expert practitioners make progress in such situations through provisional action that reveals distinctions, clarifies tractability, and sometimes reorients the direction of the work itself. Across the corpus, participants often worked before either the problem space or a direction had become intelligible enough for refinement---before a stable representational target, decision point, or viable next step had fully come into view. Under such conditions, progress depended on making the situation interpretable enough for a direction to emerge, hold, or change, alongside any revision of existing directions. These findings locate a consequential part of expert practice in the provisional action through which practitioners establish footholds, clarify tractability, and reorient the work. In this section, we position that contribution relative to existing visualization research and consider what it enables for theory, research, education, and tool support.

\subsection{Beyond Decision Points in Visualization Design}

Visualization research provides substantial support for reasoning about design. Decision-oriented models such as the Nested Model and the Blocks and Guidelines extension help articulate major design choices and their validation pressures \cite{munzner_nested_2009, meyer_nested_2015}. Process-oriented accounts such as the Design Activity Framework explicitly acknowledge that visualization work is messy, iterative, and multilinear, and connect design activities to decisions across a project \cite{mckenna_design_2014}. Design Study Methodology likewise recognizes that problem-driven visualization work may begin with fuzzy tasks and proceed through iterative, overlapping stages rather than a rigid sequence \cite{sedlmair_design_2012}. Other frameworks support reasoning about goals, tasks, activities, problems, and design spaces \cite{lam_bridging_2018, schulz_design_2013, gleicher_problem_2023, dai_how_2026, sedig_design_2016, Sedig2013}. Related methodological and empirical work has also moved toward a richer account of practice by emphasizing contextual shaping, ongoing evaluation, design rationale, and the opportunistic character of professional work \cite{McCurdy2016, wood_design_2019, hutchinson_capturing_2025, bigelow_reflections_2014, alspaugh_futzing_2019, zhang_visualization_2023, parsons_understanding_2022}. These accounts establish the messy, iterative, and multilinear character of visualization design.

Our findings suggest, however, that a major component of how design work actually proceeds remains underelaborated in existing models and frameworks. Many of the decision points they foreground are most useful when a situation is already interpretable enough for a task, abstraction, encoding, interaction, or evaluation question to be clearly posed. But in the episodes we analyzed, practitioners frequently had to determine whether the situation was primarily one of scope, structure, communication, stakeholder alignment, weak understanding, or some other unresolved condition, and what provisional move could make the work more actionable. These conditions are not merely surrounding project complications. When they shape what can count as a coherent story, a trustworthy comparison, a viable representation, or a legitimate public-facing account, they become part of the visualization problem itself. This is not a minor prelude to ``real’’ design work, nor merely a residual category of inconsequential messiness left over once the important decisions have been made. It is a consequential component of practice through which recognizable decisions and directions are often established, revised, sustained, or reopened.

This interpretation is consistent with prior empirical work showing that visualization practitioners do not report relying on systematic design processes or high-level frameworks to guide work moment by moment, but instead describe situated planning, precedent, intuition, and in-the-moment judgment as central to practice \cite{parsons_understanding_2022}. This does not mean that frameworks lack value; it suggests that an important component of practice remains underdescribed if we stop at the level of explicit decision points, process stages, or abstract task structures.

Our findings therefore complement existing models and frameworks rather than replacing them. Where those accounts help articulate decision loci, project activities, abstractions, rationales, and methods, our account makes more explicit a component of practice they often presuppose but do not fully elaborate---the situated organization of local action within unresolved design situations. Existing models are valuable for clarifying what designers may need to decide. Our findings help explain how those decision points become workable, are sustained, or are destabilized in the flow of practice. Additionally, they suggest a practical limit on what can be specified in advance. When the next meaningful move depends on what the situation reveals in response to local action, design cannot be fully guided by articulated decision structures alone.

\subsection{Interpretability as an Achievement of Design Work}

A central implication of our findings is that interpretability often emerges through visualization design rather than preceding it. Existing visualization models and frameworks are useful when a situation is interpretable enough for a clear design question to be posed---for example, what task to support, what abstraction to use, what encoding or interaction to choose, or how to evaluate a direction. Our findings suggest that this condition cannot always be assumed. In many of the episodes we analyzed, practitioners did not yet know what kind of decision was actually at stake. The issue might prove to be one of scope, structure, significance, stakeholder alignment, representational feasibility, or some other unresolved condition. What was missing was not simply the right answer, but a sufficiently settled understanding of what kind of problem or representational target the work had become. This is not merely a finer-grained description of iteration. It concerns how a situation becomes intelligible enough for recognizable design reasoning to proceed---a condition that existing visualization models often rely on but do not themselves explain.

Broader work on situated and epistemic action helps clarify why this should not be surprising. Suchman argues that action cannot be understood adequately as the execution of prior plans alone, because its organization is worked out in relation to the unfolding situation \cite{suchman_plans_1987}. Kirsh further argues that decision-cycle views of how people interact with the world are incomplete because people also act to discover possibilities, prepare and maintain environments, and make the world more cognitively tractable for ongoing reasoning \cite{kirsh_interactivity_1997, kirsh_distinguishing_1994}. In design scholarship, Schön's account of reflective conversation makes a closely related point in that designers make moves whose consequences reshape what they are able to notice and think next \cite{schon_designing_1992, schon_kinds_1992}. These traditions suggest that the issue identified in our data is not a marginal layer of messiness around otherwise orderly decision making. It is a fundamental feature of how action and understanding are coupled in open-ended design work.

The findings in Section \ref{sec:provisionalaction} make visible the importance of provisional action. Practitioners did not merely act once interpretability had been secured; they often had to act in order to produce it. Through sketching, restructuring, testing, narrowing, comparing, and trying partial representations, they clarified what the situation could support, what was tractable, and what kind of direction could hold. As Sections \ref{sec:actiondiscovery} and \ref{sec:actiontractability} show, such moves could either surface a new distinction or clarify what the work could realistically sustain. Sometimes, acting and then seeing what the situation produced was the only way to discover a path forward. This is precisely the dynamic captured in Schön's see--move--see account \cite{schon_designing_1992}: a move is made not after the situation has been fully understood, but \textit{in order to see what the situation will reveal in return}. John's account of the food-pairings visualization makes this especially clear. The distinction that ultimately organized the design did not precede action as a fully formed idea; it emerged through working the representation and reading what the situation disclosed in response. In this sense, action was part of how the situation became intelligible enough for more familiar design reasoning to proceed, not simply the execution of an already intelligible plan.

Some of the episodes resemble what is commonly described as exploratory data analysis (EDA), particularly when practitioners probe structure, test relationships, or use partial visualizations to see what the data may support. However, the point of our framing is different. Rather than treating such activity primarily as an analytic method for inspecting data, we focus on its role in design cognition---provisional action helps make an unresolved situation interpretable enough for the next representational, conceptual, or practical move to become available. In some cases, as with Omar, this involves exploring the internal structure of the data. In others, as with John, it involves working the representation itself until a more consequential distinction becomes visible. `Action as discovery' identifies how such activity functions within unresolved design work to make a situation actionable; thus, its scope extends beyond EDA.

In unresolved situations, practitioners do not operate with full visibility over the terrain of relevant possibilities. They cannot see the whole space of meaningful options, criteria, or consequences in advance. Instead, they advance through local moves whose consequences reveal only a limited portion of what the situation can support next. Design in such cases is therefore sequential and path-dependent, as each move not only advances the work, but changes what becomes visible, plausible, or tractable afterward. This differs from an idealized view of design as the enactment of a largely pre-specified sequence of moves and criteria. Our findings suggest that unresolved visualization design work more often proceeds through local disclosure, where the terrain of action becomes visible only through movement within it.

Stolterman's distinction between being \emph{prepared-for-action} and \emph{guided-in-action} sharpens this implication \cite{stolterman_nature_2008}. Repertoires, concepts, methods, and prior experience can prepare designers for unresolved work. They cannot determine the next move because what matters depends on how the particular situation responds to local action. Designers must therefore exercise judgment by reading what provisional moves disclose and proceeding accordingly. Prescriptive guidance can support this work without specifying its course in advance.

The contribution of this paper, then, is not simply to add vocabulary to familiar ideas such as iteration, exploration, or judgment. It is to make more explicit a consequential aspect of design cognition that remains underelaborated in the visualization literature---how unresolved situations become interpretable enough for recognizable design decisions to become meaningful at all, and how that interpretability can remain provisional, revisable, and vulnerable to reorientation as the work unfolds. If that is so, then situatedness is not an optional layer around design cognition. It is a necessary condition of how design cognition works in unresolved situations.

\subsection{Rethinking the Vocabulary of Visualization Design}

These findings suggest that some of the field’s familiar terms for describing visualization design are no longer sufficient. First, interpretability should not be treated primarily as a precondition for design reasoning. Much of the literature understandably begins from situations in which a task, abstraction, encoding question, or evaluation problem can already be named \cite{munzner_nested_2009, sedlmair_design_2012}. Our findings suggest that this starting point, while analytically useful, can hide an important part of practice. In unresolved situations, what designers are often working toward is not simply a solution, but a sufficiently stable understanding of what the problem, representational target, or viable direction has become. This implies that design research should be more cautious about treating clarity of problem formulation as an assumed starting condition rather than as an achievement of practice.

Second, action should not be described merely as the execution or externalization of prior reasoning. In many existing accounts, action appears after key interpretive work has already occurred---once a task has been identified, an abstraction chosen, or an encoding selected\cite{munzner_nested_2009, mckenna_design_2014, sedig_design_2016}. By contrast, our findings show that practitioners often act in order to discover what the situation can support, what is at stake, and what kind of question is becoming meaningful. This suggests that the field should place greater conceptual weight on provisional action as part of how design reasoning is constituted, not only how it is expressed, extending prior arguments that judgment in visualization practice cannot be reduced to explicit choice alone \cite{parsons_design_2020}.

Third, tractability deserves more explicit theoretical status in accounts of design practice. Visualization research often gives greatest interpretive weight to insight, creativity, or major reframings, while acts of narrowing, simplification, fallback, or feasibility-testing can appear secondary or merely practical. Instead of being peripheral, our findings suggest that such acts are among the mechanisms through which situations become workable enough for explicit representational and analytical decisions to be made. If so, then making a situation tractable should be treated not as cleanup around design work, but as part of design work's cognitive and practical core.

Finally, reorientation should not be reduced to iteration in the weak sense of repeated improvement. Iteration is an important term in visualization research \cite{mckenna_design_2014}, yet it can obscure a stronger phenomenon visible in our data. Provisional action often changes what the work is about, what counts as relevant, and what kind of representation the situation can sustain. In such cases, the issue is not simply that designers revise an existing direction, but that the direction itself is reconstituted. The language of iteration remains useful, but it is too weak on its own if it fails to distinguish refinement within an established direction from reorientation that changes what the direction is.

This perspective contributes not only another layer of descriptive nuance to existing accounts of visualization design; it also argues that the field's descriptive vocabulary must more clearly accommodate interpretability as an achievement, action as constitutive, tractability as substantive, and reorientation as generative. Without these shifts, important parts of visualization design risk being treated as residual messiness around otherwise orderly design reasoning rather than as central mechanisms through which visualization design proceeds.

\subsection{Implications for Research, Education, and Tool Support}

For research, the findings suggest that visualization design cannot be characterized only through project trajectories, stated rationales, or finished artifacts. It also requires analytic attention to bounded unresolved episodes in which designers establish footholds, make provisional moves, and reorient when the path forward is not yet clear. This shifts the target of research on design practice from design outcomes alone to the situated processes through which outcomes become possible. It also implies that design theories should be judged not only by how well they articulate explicit decisions, but also by how well they explain how those decisions become meaningful in the first place.

For education, the findings suggest that learning design methods is not enough. Students may find value in frameworks for tasks, abstractions, encodings, iteration, and evaluation. But many of the difficulties novices face arise earlier and more locally than those categories imply. They do not know what deserves attention, what kind of move to make, what to simplify, when to narrow scope, or when an apparent direction is not holding. If interpretability is often achieved rather than given, then pedagogy must prepare students to proceed under partial intelligibility. In Stolterman's terms, education can prepare students for action through repertoires, concepts, and methods, but it cannot fully guide them in action through every unresolved situation. This means helping students learn how to establish footholds, make provisional moves, recognize tractability, and understand reorientation as a normal consequence of open-ended work.

For tool support, the findings suggest that current systems may privilege decision-ready states and under-serve unresolved work. Many visualization tools, including emerging AI-assisted systems, are optimized for generating candidate charts, recommending encodings, translating specifications, or evaluating alternatives once a direction has become sufficiently articulated. Recent work on visualization design knowledge helps explain this tendency by distinguishing forms of design knowledge that can be codified from knowledge-in-use that depends on situated interpretation and judgment \cite{parsons_design_2027}. Our findings suggest that support is also needed earlier and more locally, including for restructuring data to see what holds, trying partial views, surfacing the consequences of tentative moves, comparing possible footholds, and noticing when a problem formulation or representation is beginning to fail. Tools that assume a stable question too early may encourage premature closure. Tools can support unresolved situations by preparing designers for action while leaving the course of action open. Systems that facilitate provisional moves and make their consequences visible may therefore align more closely with how expert practice proceeds under unresolved conditions.

\subsection{Limitations}

This study has several limitations that clarify the scope of the contribution. First, the analysis is based on a relatively small corpus involving eleven expert practitioners. Our aim was analytic depth and conceptual development rather than statistical generalization, so the findings should be understood as recurring patterns in this corpus rather than as an exhaustive characterization of visualization practice. Second, although the three-phase design includes challenge materials, in-flight diary entries, interviews, and associated artifacts, some of the material remains retrospective in character. The corpus therefore provides strong access to participants' accounts of unresolved episodes, but not continuous observational access to action as it unfolded in situ. Third, the study centers primarily on individual practitioners. While participants discussed collaborators, clients, and organizational constraints, the analysis does not directly examine unresolved situations as distributed across teams or institutions. Fourth, the paper focuses on expert practice and does not compare experts with novices or less experienced practitioners. The findings therefore clarify a consequential component of expert work, but do not by themselves explain how these capacities develop across levels of experience.

\subsection{Future Research Directions}

Future work should study unresolved episodes more directly in naturalistic settings through richer observational methods, providing finer-grained access to how provisional action unfolds in real time and how local consequences become visible, negotiated, and acted upon. It should also extend this analysis beyond individual practitioners to collaborative and organizational contexts, examining how unresolved situations are jointly interpreted and how tractability and reorientation are distributed across teams, clients, and institutions. Because our study focuses on expert practitioners, future work could compare experts and novices to investigate how foothold-building, tractability recognition, and reorientation develop with experience, and how these capacities might be better supported in education. The findings also motivate closer study of tool support, especially AI-assisted design systems, including which kinds of systems help designers remain productively engaged with partial intelligibility and which instead encourage premature closure by over-privileging decision-ready states.

\section{Conclusion}

This paper has presented an empirical account of how unresolved situations become actionable in visualization design practice. Through an analysis of a multi-phase qualitative corpus involving eleven expert practitioners, we showed that important moments of design work do not always begin from a clearly interpretable problem, a stable representational target, or a well-defined next step. Instead, practitioners make provisional moves that reveal what the situation can support, what is tractable, and when the work requires reorientation. Existing visualization frameworks remain valuable for articulating decision loci, activities, abstractions, and rationales, but our findings suggest that they are often strongest once the situation has become interpretable enough for those decisions to be meaningfully posed. Our contribution is to make more explicit the situated organization of action through which that interpretability is achieved, and with it a practical limit on what can be specified in advance. When the next meaningful move depends on what the situation reveals in response to local action, design cannot be fully guided by prescriptive decision structures alone.

\section*{Supplemental Materials}
\label{sec:supplemental_materials}
Supplementary materials provide additional detail on the analysis reported in this paper. They include a reconstructed episode-level inventory and expanded illustrative cases that show how the reported patterns were grounded in the corpus.

%% if specified like this the section will be omitted in review mode
\acknowledgments{%
This work was supported in part by a grant from the NSF (\# 2146228).
}

\bibliographystyle{abbrv-doi-hyperref}

\bibliography{references}

@incollection{Kirsh2009,
	title = {Problem solving and situated cognition},
	doi = {10.1017/CBO9780511816826.015},
	urldate = {2011-01-04},
	booktitle = {The {Cambridge} handbook of situated cognition},
	publisher = {Cambridge University Press},
	author = {Kirsh, David},
	editor = {Robbins, Philip and Aydede, Murat},
	year = {2009},
	pages = {264--306},
}

@article{zhang_visualization_2023,
	title = {Visualization {Design} {Practices} in a {Crisis}: {Behind} the {Scenes} {With} {COVID}-19 {Dashboard} {Creators}},
	volume = {29},
	copyright = {https://ieeexplore.ieee.org/Xplorehelp/downloads/license-information/IEEE.html},
	issn = {1077-2626, 1941-0506, 2160-9306},
	shorttitle = {Visualization {Design} {Practices} in a {Crisis}},
	url = {https://ieeexplore.ieee.org/document/9904866/},
	doi = {10.1109/TVCG.2022.3209493},
	language = {en},
	number = {1},
	urldate = {2025-04-16},
	journal = {IEEE Transactions on Visualization and Computer Graphics},
	author = {Zhang, Yixuan and Sun, Yifan and Gaggiano, Joseph D. and Kumar, Neha and Andris, Clio and Parker, Andrea G.},
	year = {2023},
	pages = {1037--1047},
}

@article{parsons_understanding_2022,
	title = {Understanding {Data} {Visualization} {Design} {Practice}},
	volume = {28},
	copyright = {All rights reserved},
	doi = {10.1109/TVCG.2021.3114959},
	number = {1},
	journal = {IEEE Transactions on Visualization and Computer Graphics},
	author = {Parsons, Paul},
	year = {2022},
	pages = {665--675},
}

@article{lee-robbins_client-designer_2025,
	title = {Client-{Designer} {Negotiation} in {Data} {Visualization} {Projects}},
	volume = {31},
	copyright = {https://ieeexplore.ieee.org/Xplorehelp/downloads/license-information/IEEE.html},
	issn = {1077-2626, 1941-0506, 2160-9306},
	url = {https://ieeexplore.ieee.org/document/10704568/},
	doi = {10.1109/TVCG.2024.3467189},
	language = {en},
	number = {9},
	urldate = {2025-03-20},
	journal = {IEEE Transactions on Visualization and Computer Graphics},
	author = {Lee-Robbins, Elsie and Ridley, Arran and Adar, Eytan},
	year = {2025},
	pages = {5772--5787},
}

@article{bako_understanding_2023,
	title = {Understanding {How} {Designers} {Find} and {Use} {Data} {Visualization} {Examples}},
	volume = {29},
	issn = {1077-2626, 1941-0506, 2160-9306},
	url = {https://ieeexplore.ieee.org/document/9903579/},
	doi = {10.1109/TVCG.2022.3209490},
	language = {en},
	number = {1},
	urldate = {2023-06-22},
	journal = {IEEE Transactions on Visualization and Computer Graphics},
	author = {Bako, Hannah K. and Liu, Xinyi and Battle, Leilani and Liu, Zhicheng},
	year = {2023},
	pages = {1048--1058},
}

@inproceedings{parsons_judgment_2025,
	address = {Vienna, Austria},
	title = {Judgment as {Coordination}: {A} {Joint} {Systems} {View} of {Visualization} {Design} {Practice}},
	shorttitle = {Judgment as {Coordination}},
	doi = {10.1109/VIS60296.2025.00062},
	language = {en},
	urldate = {2025-08-08},
	booktitle = {{IEEE} {Conference} on {Visualization} and {Visual} {Analytics} ({VIS})},
	author = {Parsons, Paul C. and Ridley, Arran},
	year = {2025},
	pages = {281--285},
}

@article{munzner_nested_2009,
	title = {A nested model for visualization design and validation},
	volume = {15},
	doi = {10.1109/TVCG.2009.111},
	number = {6},
	journal = {IEEE Transactions on Visualization and Computer Graphics},
	author = {Munzner, Tamara},
	year = {2009},
	pages = {921--928},
}

@article{schon_designing_1992,
	title = {Designing as a {Reflective} {Conversation} with the {Materials} of a {Design} {Situation}},
	volume = {3},
	doi = {10.1007/BF01580516},
	urldate = {2021-01-15},
	journal = {Research in Engineering Design},
	author = {Schön, Donald A.},
	year = {1992},
	pages = {131--147},
}

@article{parsons_design_2027,
	title = {Design {Knowledge} in {Data} {Visualization}: {Mapping} the {Epistemic} {Landscape}},
	language = {en},
	journal = {IEEE Transactions on Visualization and Computer Graphics},
	author = {Parsons, Paul C and Gray, Colin M and Baigelenov, Ali},
	year = {2027},
	note = {In Press},
}

@misc{nightingale_editors_data_2022,
	title = {Data {Is} {Plural} {Submissions}: {London} {Stage} {Database}},
	url = {https://nightingaledvs.com/data-is-plural-submissions-london-stage-database/},
	urldate = {2023-09-17},
	author = {{Nightingale Editors}},
	month = dec,
	year = {2022},
	note = {https://nightingaledvs.com/data-is-plural-submissions-london-stage-database/},
}

@article{mcdonald_reliability_2019,
	title = {Reliability and {Inter}-rater {Reliability} in {Qualitative} {Research}: {Norms} and {Guidelines} for {CSCW} and {HCI} {Practice}},
	volume = {3},
	doi = {10.1145/3359174},
	language = {en},
	journal = {Proceedings of the ACM on Human-Computer Interaction},
	author = {McDonald, Nora and Schoenebeck, Sarita and Forte, Andrea},
	year = {2019},
	pages = {1--23},
}

@article{fereday_demonstrating_2006,
	title = {Demonstrating rigor using thematic analysis: {A} hybrid approach of inductive and deductive coding and theme development},
	volume = {5},
	doi = {10.1177/160940690600500107},
	number = {1},
	journal = {International Journal of Qualitative Methods},
	author = {Fereday, Jennifer and Muir-Cochrane, Eimear},
	year = {2006},
	pages = {80--92},
}

@article{Cross1981,
	title = {Design method and scientific method},
	volume = {2},
	url = {http://www.sciencedirect.com/science/article/pii/0142694X81900508},
	doi = {10.1016/0142-694X(81)90050-8},
	number = {4},
	urldate = {2012-08-05},
	journal = {Design studies},
	author = {Cross, Nigel and Naughton, J. and Walker, D.},
	year = {1981},
	pages = {195--201},
}

@article{buchanan_wicked_1992,
	title = {Wicked {Problems} in {Design} {Thinking}},
	volume = {8},
	doi = {10.2307/1511637},
	number = {2},
	journal = {Design Issues},
	author = {Buchanan, Richard},
	year = {1992},
	pages = {5--21},
}

@article{schon_designing_1988,
	title = {Designing: {Rules}, types and worlds},
	volume = {9},
	doi = {10.1016/0142-694X(88)90047-6},
	language = {en},
	number = {3},
	journal = {Design Studies},
	author = {Schön, Donald A},
	year = {1988},
	pages = {181--190},
}

@book{cross_design_2011,
	title = {Design {Thinking}: {Understanding} {How} {Designers} {Think} and {Work}},
	isbn = {978-1-84788-846-4},
	shorttitle = {Design {Thinking}},
	language = {en},
	publisher = {Berg},
	author = {Cross, Nigel},
	month = apr,
	year = {2011},
}

@book{Dorst_2015_FrameInnovation,
	series = {Design {Thinking}},
	title = {Frame {Innovation}: {Create} {New} {Thinking} by {Design}},
	isbn = {978-0-262-32885-2},
	shorttitle = {Frame {Innovation}},
	language = {en},
	publisher = {MIT Press},
	author = {Dorst, Kees},
	month = apr,
	year = {2015},
}

@article{kindlmann_algebraic_2014,
	title = {An {Algebraic} {Process} for {Visualization} {Design}},
	volume = {20},
	copyright = {https://ieeexplore.ieee.org/Xplorehelp/downloads/license-information/IEEE.html},
	issn = {1077-2626},
	url = {http://ieeexplore.ieee.org/document/6875930/},
	doi = {10.1109/TVCG.2014.2346325},
	language = {en},
	number = {12},
	urldate = {2026-03-28},
	journal = {IEEE Transactions on Visualization and Computer Graphics},
	author = {Kindlmann, Gordon and Scheidegger, Carlos},
	month = dec,
	year = {2014},
	pages = {2181--2190},
}

@article{wood_design_2019,
	title = {Design {Exposition} with {Literate} {Visualization}},
	volume = {25},
	copyright = {https://ieeexplore.ieee.org/Xplorehelp/downloads/license-information/IEEE.html},
	issn = {1077-2626, 1941-0506, 2160-9306},
	url = {https://ieeexplore.ieee.org/document/8440080/},
	doi = {10.1109/TVCG.2018.2864836},
	language = {en},
	number = {1},
	urldate = {2026-03-28},
	journal = {IEEE Transactions on Visualization and Computer Graphics},
	author = {Wood, Jo and Kachkaev, Alexander and Dykes, Jason},
	month = jan,
	year = {2019},
	pages = {759--768},
}

@inproceedings{hutchinson_capturing_2025,
	address = {Vienna, Austria},
	title = {Capturing {Visualization} {Design} {Rationale}},
	copyright = {https://doi.org/10.15223/policy-029},
	isbn = {979-8-3315-6613-5},
	url = {https://ieeexplore.ieee.org/document/11298820/},
	doi = {10.1109/VIS60296.2025.00052},
	language = {en},
	urldate = {2026-03-28},
	booktitle = {2025 {IEEE} {Visualization} and {Visual} {Analytics} ({VIS})},
	publisher = {IEEE},
	author = {Hutchinson, Maeve and Jianu, Radu and Slingsby, Aidan and Wood, Jo and Madhyastha, Pranava},
	month = nov,
	year = {2025},
	pages = {231--235},
}

@article{gleicher_problem_2023,
	title = {A {Problem} {Space} for {Designing} {Visualizations}},
	volume = {43},
	issn = {1558-1756},
	url = {https://ieeexplore.ieee.org/abstract/document/10179119},
	doi = {10.1109/MCG.2023.3267213},
	number = {4},
	urldate = {2026-03-22},
	journal = {IEEE Computer Graphics and Applications},
	author = {Gleicher, Michael and Riveiro, Maria and von Landesberger, Tatiana and Deussen, Oliver and Chang, Remco and Gillman, Christina},
	month = jul,
	year = {2023},
	pages = {111--120},
}

@article{dai_how_2026,
	title = {How {We} {Map} {Possibilities}: {Understanding} {Design} {Spaces} for {Visualization}},
	issn = {1941-0506},
	shorttitle = {How {We} {Map} {Possibilities}},
	url = {https://ieeexplore.ieee.org/abstract/document/11442634},
	doi = {10.1109/TVCG.2026.3675300},
	urldate = {2026-03-22},
	journal = {IEEE Transactions on Visualization and Computer Graphics},
	author = {Dai, Zichun and Peng, Yechun and Cao, Nan and Shi, Yang},
	year = {2026},
	pages = {1--15},
}

@article{stolterman_nature_2008,
	title = {The {Nature} of {Design} {Practice} and {Implications} for {Interaction} {Design} {Research}},
	volume = {2},
	issn = {09501991},
	number = {1},
	journal = {International Journal of Design},
	author = {Stolterman, Erik},
	year = {2008},
	pages = {55--65},
}

@article{schon_problems_1984,
	title = {Problems, frames and perspectives on designing},
	volume = {5},
	issn = {0142-694X},
	url = {https://www.sciencedirect.com/science/article/pii/0142694X84900024},
	doi = {10.1016/0142-694X(84)90002-4},
	number = {3},
	journal = {Design Studies},
	author = {Schön, Donald A.},
	year = {1984},
	pages = {132--136},
}

@article{parsons_beyond_2026,
	title = {Beyond {Problem} {Solving}: {Framing} and {Problem}–{Solution} {Co}-{Evolution} in {Data} {Visualization} {Design}},
	volume = {32},
	doi = {10.1109/TVCG.2025.3633866},
	language = {en},
	number = {1},
	journal = {IEEE Transactions on Visualization and Computer Graphics},
	author = {Parsons, Paul C and Shukla, Prakash Chandra},
	year = {2026},
	pages = {24--34},
}

@article{christensen_kinds_2021,
	title = {Kinds of ‘moving’ in designing with sticky notes},
	volume = {76},
	issn = {0142694X},
	url = {https://linkinghub.elsevier.com/retrieve/pii/S0142694X21000478},
	doi = {10.1016/j.destud.2021.101036},
	language = {en},
	urldate = {2026-03-13},
	journal = {Design Studies},
	author = {Christensen, Bo T. and Abildgaard, Sille Julie J.},
	month = sep,
	year = {2021},
	pages = {101036},
}

@article{schulz_design_2013,
	title = {A {Design} {Space} of {Visualization} {Tasks}},
	volume = {19},
	issn = {1941-0506},
	url = {https://ieeexplore.ieee.org/abstract/document/6634156},
	doi = {10.1109/TVCG.2013.120},
	number = {12},
	urldate = {2026-03-09},
	journal = {IEEE Transactions on Visualization and Computer Graphics},
	author = {Schulz, Hans-Jörg and Nocke, Thomas and Heitzler, Magnus and Schumann, Heidrun},
	month = dec,
	year = {2013},
	pages = {2366--2375},
}

@article{schon_kinds_1992,
	title = {Kinds of seeing and their functions in designing},
	volume = {13},
	copyright = {https://www.elsevier.com/tdm/userlicense/1.0/},
	issn = {0142694X},
	url = {https://linkinghub.elsevier.com/retrieve/pii/0142694X9290268F},
	doi = {10.1016/0142-694X(92)90268-F},
	language = {en},
	number = {2},
	urldate = {2025-09-11},
	journal = {Design Studies},
	author = {Schön, Donald A. and Wiggins, Glenn},
	month = apr,
	year = {1992},
	pages = {135--156},
}

@article{atman_engineering_2007,
	title = {Engineering design processes: {A} comparison of students and expert practitioners},
	volume = {96},
	doi = {10.1002/j.2168-9830.2007.tb00945.x},
	number = {4},
	journal = {Journal of Engineering Education},
	author = {Atman, Cynthia J and Adams, Robin S and Cardella, Monica E and Turns, Jennifer and Mosborg, Susan and Saleem, Jason},
	year = {2007},
	pages = {359--379},
}

@article{hoffman_identifying_2023,
	title = {Identifying {Experts} for the {Design} of {Human}-{Centered} {Systems}: {The} {Pentapod} {Principle}},
	volume = {6},
	language = {en},
	number = {3},
	journal = {Journal of Expertise},
	author = {Hoffman, Robert R},
	year = {2023},
	pages = {259--266},
}

@article{rittel_dilemmas_1973,
	title = {Dilemmas in a general theory of planning},
	volume = {4},
	doi = {10.1007/BF01405730},
	language = {en},
	journal = {Policy Sciences},
	author = {Rittel, Horst W. J. and Webber, Melvin M.},
	year = {1973},
	pages = {155--169},
}

@inproceedings{parsons_design_2020,
	address = {Salt Lake City, UT},
	title = {Design {Judgment} in {Data} {Visualization} {Practice}},
	doi = {10.1109/VIS47514.2020.00042},
	language = {en},
	urldate = {2021-02-17},
	booktitle = {{IEEE} {Visualization} {Conference} ({VIS})},
	author = {Parsons, Paul and Gray, Colin M. and Baigelenov, Ali and Carr, Ian},
	month = sep,
	year = {2020},
	pages = {176--180},
}

@inproceedings{parsons_fixation_2021,
	address = {New Orleans, LA},
	title = {Fixation and {Creativity} in {Data} {Visualization} {Design}: {Experiences} and {Perspectives} of {Practitioners}},
	copyright = {All rights reserved},
	doi = {10.1109/VIS49827.2021.9623297},
	booktitle = {{IEEE} {Visualization} {Conference} ({VIS})},
	author = {Parsons, Paul and Shukla, Prakash Chandra and Park, Chorong},
	year = {2021},
	pages = {76--80},
}

@phdthesis{ridley_sociocultural_2021,
	type = {{PhD} {Thesis}},
	title = {A sociocultural approach to data visualisation evaluation: understanding the role of a large-scale tool-based approach to evaluation within the commercial production process.},
	school = {University of Leeds},
	author = {Ridley, Arran Lee},
	year = {2021},
}

@inproceedings{baigelenov_how_2025,
	address = {New York, NY, USA},
	series = {{CHI} '25},
	title = {How {Visualization} {Designers} {Perceive} and {Use} {Inspiration}},
	isbn = {979-8-4007-1394-1},
	url = {https://dl.acm.org/doi/10.1145/3706598.3714191},
	doi = {10.1145/3706598.3714191},
	urldate = {2025-04-29},
	booktitle = {Proceedings of the 2025 {CHI} {Conference} on {Human} {Factors} in {Computing} {Systems}},
	publisher = {Association for Computing Machinery},
	author = {Baigelenov, Ali and Shukla, Prakash and Parsons, Paul},
	month = apr,
	year = {2025},
	pages = {1--13},
}

@incollection{swain_hybrid_2018,
	series = {Cases},
	title = {A hybrid approach to thematic analysis in qualitative research: {Using} a practical example},
	volume = {Part 2},
	shorttitle = {A hybrid approach to thematic analysis in qualitative research},
	booktitle = {Sage {Research} {Methods}},
	publisher = {SAGE Publications, Ltd.},
	author = {Swain, Jon},
	year = {2018},
}

@inproceedings{williams_data_2023,
	address = {Hamburg Germany},
	title = {Data {Abstraction} {Elephants}: {The} {Initial} {Diversity} of {Data} {Representations} and {Mental} {Models}},
	isbn = {978-1-4503-9421-5},
	shorttitle = {Data {Abstraction} {Elephants}},
	url = {https://dl.acm.org/doi/10.1145/3544548.3580669},
	doi = {10.1145/3544548.3580669},
	language = {en},
	urldate = {2025-03-20},
	booktitle = {Proceedings of the 2023 {CHI} {Conference} on {Human} {Factors} in {Computing} {Systems}},
	publisher = {ACM},
	author = {Williams, Katy and Bigelow, Alex and Isaacs, Katherine E.},
	month = apr,
	year = {2023},
	pages = {1--24},
}

@article{maher_co-evolution_2003,
	title = {Co-evolution as a computational and cognitive model of design},
	volume = {14},
	copyright = {http://www.springer.com/tdm},
	issn = {0934-9839},
	url = {http://link.springer.com/10.1007/s00163-002-0016-y},
	doi = {10.1007/s00163-002-0016-y},
	language = {en},
	number = {1},
	urldate = {2024-05-23},
	journal = {Research in Engineering Design},
	author = {Maher, Mary and Tang, Hsien-Hui},
	month = feb,
	year = {2003},
	pages = {47--64},
}

@article{lam_bridging_2018,
	title = {Bridging from {Goals} to {Tasks} with {Design} {Study} {Analysis} {Reports}},
	volume = {24},
	issn = {1077-2626},
	url = {http://ieeexplore.ieee.org/document/8023762/},
	doi = {10.1109/TVCG.2017.2744319},
	language = {en},
	number = {1},
	urldate = {2023-02-28},
	journal = {IEEE Transactions on Visualization and Computer Graphics},
	author = {Lam, Heidi and Tory, Melanie and Munzner, Tamara},
	month = jan,
	year = {2018},
	pages = {435--445},
}

@article{hay_future_2020,
	title = {The future of design cognition analysis},
	volume = {6},
	issn = {2053-4701},
	url = {https://www.cambridge.org/core/product/identifier/S2053470120000207/type/journal_article},
	doi = {10.1017/dsj.2020.20},
	language = {en},
	urldate = {2022-09-16},
	journal = {Design Science},
	author = {Hay, Laura and Cash, Philip and McKilligan, Seda},
	year = {2020},
	pages = {e20},
}

@article{alspaugh_futzing_2019,
	title = {Futzing and {Moseying}: {Interviews} with {Professional} {Data} {Analysts} on {Exploration} {Practices}},
	volume = {25},
	issn = {1077-2626, 1941-0506, 2160-9306},
	shorttitle = {Futzing and {Moseying}},
	url = {https://ieeexplore.ieee.org/document/8440815/},
	doi = {10.1109/TVCG.2018.2865040},
	language = {en},
	number = {1},
	urldate = {2021-07-20},
	journal = {IEEE Transactions on Visualization and Computer Graphics},
	author = {Alspaugh, Sara and Zokaei, Nava and Liu, Andrea and Jin, Cindy and Hearst, Marti A.},
	month = jan,
	year = {2019},
	pages = {22--31},
}

@article{kirsh_distinguishing_1994,
	title = {On {Distinguishing} {Epistemic} from {Pragmatic} {Action}},
	volume = {18},
	issn = {03640213},
	url = {http://www.informaworld.com/openurl?genre=article&doi=10.1207/s15516709cog1804_1&magic=crossref%7C%7CD404A21C5BB053405B1A640AFFD44AE3},
	doi = {10.1207/s15516709cog1804_1},
	number = {4},
	journal = {Cognitive Science},
	author = {Kirsh, David and Maglio, Paul},
	month = oct,
	year = {1994},
	pages = {513--549},
}

@article{kirsh_interactivity_1997,
	title = {Interactivity and multimedia interfaces},
	volume = {25},
	issn = {00204277},
	doi = {10.1023/A:1002915430871},
	number = {2},
	urldate = {2010-08-22},
	journal = {Instructional Science},
	author = {Kirsh, David},
	year = {1997},
	pages = {79--96},
}

@article{walny_data_2020,
	title = {Data {Changes} {Everything}: {Challenges} and {Opportunities} in {Data} {Visualization} {Design} {Handoff}},
	volume = {26},
	issn = {1077-2626, 1941-0506, 2160-9306},
	shorttitle = {Data {Changes} {Everything}},
	url = {https://ieeexplore.ieee.org/document/8816695/},
	doi = {10.1109/TVCG.2019.2934538},
	language = {en},
	number = {1},
	urldate = {2020-10-26},
	journal = {IEEE Transactions on Visualization and Computer Graphics},
	author = {Walny, Jagoda and Frisson, Christian and West, Mieka and Kosminsky, Doris and Knudsen, Soren and Carpendale, Sheelagh and Willett, Wesley},
	month = jan,
	year = {2020},
	pages = {12--22},
}

@article{bigelow_iterating_2017,
	title = {Iterating between {Tools} to {Create} and {Edit} {Visualizations}},
	volume = {23},
	issn = {1077-2626},
	url = {http://ieeexplore.ieee.org/document/7539580/},
	doi = {10.1109/TVCG.2016.2598609},
	language = {en},
	number = {1},
	urldate = {2020-08-24},
	journal = {IEEE Transactions on Visualization and Computer Graphics},
	author = {Bigelow, Alex and Drucker, Steven and Fisher, Danyel and Meyer, Miriah},
	month = jan,
	year = {2017},
	pages = {481--490},
}

@book{suchman_plans_1987,
	title = {Plans and {Situated} {Actions}: {The} {Problem} of {Human}-{Machine} {Communication}},
	isbn = {978-0-521-33739-7},
	shorttitle = {Plans and {Situated} {Actions}},
	language = {en},
	publisher = {Cambridge University Press},
	author = {Suchman, Lucy A.},
	year = {1987},
}

@article{sedlmair_design_2012,
	title = {Design {Study} {Methodology}: {Reflections} from the {Trenches} and the {Stacks}},
	volume = {18},
	issn = {1077-2626},
	shorttitle = {Design {Study} {Methodology}},
	doi = {10.1109/TVCG.2012.213},
	number = {12},
	journal = {IEEE Transactions on Visualization and Computer Graphics},
	author = {Sedlmair, Michael and Meyer, Miriah and Munzner, Tamara},
	year = {2012},
	pages = {2431--2440},
}

@article{meyer_nested_2015,
	title = {The nested blocks and guidelines model},
	volume = {14},
	issn = {1473-8716, 1473-8724},
	url = {http://journals.sagepub.com/doi/10.1177/1473871613510429},
	doi = {10.1177/1473871613510429},
	language = {en},
	number = {3},
	urldate = {2018-06-15},
	journal = {Information Visualization},
	author = {Meyer, Miriah and Sedlmair, Michael and Quinan, P Samuel and Munzner, Tamara},
	year = {2015},
	pages = {234--249},
}

@article{mckenna_design_2014,
	title = {Design {Activity} {Framework} for {Visualization} {Design}},
	volume = {20},
	issn = {1077-2626},
	doi = {10.1109/TVCG.2014.2346331},
	number = {12},
	journal = {IEEE Transactions on Visualization and Computer Graphics},
	author = {McKenna, S. and Mazur, D. and Agutter, J. and Meyer, M.},
	year = {2014},
	pages = {2191--2200},
}

@inproceedings{McCurdy2016,
	address = {New York, New York, USA},
	title = {Action {Design} {Research} and {Visualization} {Design}},
	isbn = {978-1-4503-4818-8},
	url = {http://dl.acm.org/citation.cfm?doid=2993901.2993916},
	doi = {10.1145/2993901.2993916},
	booktitle = {Proceedings of the 6th {Biannual} {Workshop} on evaluation and {BEyond} - {methodoLogIcal} approaches for {Visualization} ({BELIV})},
	publisher = {ACM Press},
	author = {McCurdy, Nina and Dykes, Jason and Meyer, Miriah},
	year = {2016},
	pages = {10--18},
}

@inproceedings{bigelow_reflections_2014,
	address = {Como, Italy},
	title = {Reflections on how designers design with data},
	isbn = {978-1-4503-2775-6},
	url = {http://dl.acm.org/citation.cfm?doid=2598153.2598175},
	doi = {10.1145/2598153.2598175},
	language = {en},
	urldate = {2019-07-17},
	booktitle = {Proceedings of the 2014 {International} {Working} {Conference} on {Advanced} {Visual} {Interfaces} - {AVI} '14},
	publisher = {ACM Press},
	author = {Bigelow, Alex and Drucker, Steven and Fisher, Danyel and Meyer, Miriah},
	year = {2014},
	pages = {17--24},
}

@book{schon_reflective_1983,
	title = {The {Reflective} {Practitioner}: {How} {Professionals} {Think} {In} {Action}},
	publisher = {Basic Books},
	author = {Schön, Donald A.},
	year = {1983},
}

@article{braun_using_2006,
	title = {Using thematic analysis in psychology},
	volume = {3},
	issn = {1478-0887, 1478-0895},
	url = {http://www.tandfonline.com/doi/abs/10.1191/1478088706qp063oa},
	doi = {10.1191/1478088706qp063oa},
	language = {en},
	number = {2},
	urldate = {2018-11-22},
	journal = {Qualitative Research in Psychology},
	author = {Braun, Virginia and Clarke, Victoria},
	month = jan,
	year = {2006},
	pages = {77--101},
}

@article{brehmer_multi-level_2013,
	title = {A multi-level typology of abstract visualization tasks.},
	volume = {19},
	issn = {1941-0506},
	shorttitle = {Visualization and {Computer} {Graphics}, {IEEE} {Transact}},
	url = {http://www.ncbi.nlm.nih.gov/pubmed/24051804},
	doi = {10.1109/TVCG.2013.124},
	number = {12},
	urldate = {2015-07-30},
	journal = {IEEE transactions on visualization and computer graphics},
	author = {Brehmer, Matthew and Munzner, Tamara},
	month = dec,
	year = {2013},
	pages = {2376--85},
}

@book{sedig_design_2016,
	series = {Synthesis {Lectures} on {Visualization}},
	title = {Design of {Visualizations} for {Human}-{Information} {Interaction}: {A} {Pattern}-{Based} {Framework}},
	volume = {4},
	copyright = {All rights reserved},
	issn = {2159-516X},
	url = {http://www.morganclaypool.com/doi/abs/10.2200/S00685ED1V01Y201512VIS005},
	doi = {10.2200/S00685ED1V01Y201512VIS005},
	language = {en},
	urldate = {2016-04-24},
	publisher = {Morgan \& Claypool Publishers},
	author = {Sedig, Kamran and Parsons, Paul},
	editor = {Ebert, David and Elmqvist, Niklas},
	month = apr,
	year = {2016},
}

@article{chiu_design_2003,
	title = {Design moves in situated design with case-based reasoning},
	volume = {24},
	issn = {0142694X},
	url = {http://linkinghub.elsevier.com/retrieve/pii/S0142694X02000078},
	doi = {10.1016/S0142-694X(02)00007-8},
	language = {en},
	number = {1},
	urldate = {2018-09-21},
	journal = {Design Studies},
	author = {Chiu, Mao-Lin},
	month = jan,
	year = {2003},
	pages = {1--25},
}

@article{Sedig2013,
	title = {Interaction {Design} for {Complex} {Cognitive} {Activities} with {Visual} {Representations}: {A} {Pattern}-{Based} {Approach}},
	volume = {5},
	url = {http://aisel.aisnet.org/cgi/viewcontent.cgi?article=1057&context=thci},
	number = {2},
	journal = {AIS Transactions on Human-Computer Interaction},
	author = {Sedig, Kamran and Parsons, Paul},
	year = {2013},
	pages = {84--133},
}

\end{document}